\documentclass[ aip, rsi, amsmath,amssymb, reprint, ]{revtex4-1}
\usepackage{graphicx}
\usepackage{dcolumn}
\usepackage{bm}
\usepackage{bbm}
\usepackage[usenames]{color}
\begin{document}

 
\title{Time-delayed feedback control of coherence resonance chimeras
}

\author{Anna Zakharova}
\email{anna.zakharova@tu-berlin.de}
\affiliation{ Institut f\"ur Theoretische Physik, Technische Universit\"at Berlin, Hardenbergstr. 36, 10623 Berlin, Germany}
\author{Nadezhda Semenova}
\affiliation{ Department of Physics, Saratov State University, Astrakhanskaya str. 83, 410012 Saratov, Russia}
\author{Vadim Anishchenko}
\affiliation{ Department of Physics, Saratov State University, Astrakhanskaya str. 83, 410012 Saratov, Russia}
\author{Eckehard Sch\"oll}
\affiliation{ Institut f\"ur Theoretische Physik, Technische Universit\"at Berlin, Hardenbergstr. 36, 10623 Berlin, Germany}

\date{\today}

\begin{abstract}
Using the model of a FitzHugh-Nagumo system in the excitable regime we investigate the influence of time-delayed feedback on noise-induced chimera states in a network with nonlocal coupling, i.e., coherence resonance chimeras. It is shown that time-delayed feedback allows for control of the range of parameter values where these chimera states occur. Moreover, for the feedback delay close to the intrinsic period of the system we find a novel regime which we call period-two coherence resonance chimera.
\end{abstract}


\maketitle

\begin{quotation}
Coherence resonance chimeras in nonlocally coupled networks of excitable elements represent partial synchronization patterns composed of spatially separated domains of coherent and incoherent spiking behavior, which are induced by noise. These patterns are different from classical chimera states occurring in deterministic oscillatory systems and combine properties of the counter-intuitive phenomenon of coherence resonance, i.e., a constructive role of noise, and chimera states, i.e., the coexistence of spatially synchronized and desynchronized domains in a network of identical elements. Another distinctive feature of the particular type of chimera we study here is its alternating behavior, i.e., periodic switching of the location of coherent and incoherent domains.
Applying time-delayed feedback, we demonstrate how to control coherence resonance chimeras by adjusting delay time and feedback strength. In particular, we show that feedback increases the parameter intervals of existence of chimera states and has a significant impact on their alternating dynamics leading to the appearance of novel patterns, which we call period-two coherence resonance chimera. Since the dynamics of every individual network element in our study is given by the FitzHugh-Nagumo system, which is a paradigmatic model for neurons in the excitable regime, we expect wide-range applications of our results to neural networks.

\end{quotation}


\section{Introduction}\label{sec:intro}
The processes occurring in nature are inevitably affected by internal and external random fluctuations, i.e., noise. Even at a relatively low intensity, noise can significantly influence the behavior of a dynamical system. Noise can play a constructive role and give rise to new dynamic behavior, e.g., stochastic bifurcations, stochastic synchronization, or
coherence resonance \cite{HU93a,PIK97,NEI97,LIN04,USH05,ZAK10a,ZAK13}. The counter-intuitive effect of coherence resonance describes a non-monotonic behavior of the regularity
of noise-induced oscillations in the excitable regime, leading to an optimum response in terms of regularity of the excited oscillations for an intermediate noise strength. It has been previously shown that coherence resonance can be modulated by applying time-delayed feedback in excitable \cite{JAN04,AUS09} as well as in non-excitable systems \cite{GEF14,SEM15}.  In particular, for appropriate choices of time delay, either suppression or enhancement of coherence resonance can be achieved.

Recently, a new type of coherence resonance, {\em coherence resonance chimeras}, has been discovered \cite{SEM16,ZAK17}. It combines temporal features of coherence resonance, and spatial properties of chimera states \cite{PAN15,SCH16b}, i.e., coexistence of spatially coherent and incoherent domains in a network of identical elements. This phenomenon is distinct from classical chimeras, which occur in deterministic oscillatory elements \cite{KUR02a,ABR04}. It is well-known that in the presence of time delay simple dynamical systems can exhibit complex behavior, such as delay-induced bifurcations \cite{SCH03h}, delay-induced multistability \cite{HIZ07}, stabilization of unstable periodic orbits \cite{PYR92} or stationary states \cite{HOE05}, to name just a few examples. Chimera states have been investigated for noisy systems \cite{LOO16} and delayed systems as well. In general chimera patterns tend to form clusters in the presence of time delay \cite{SET08,SEN10a}. The role of time-delayed coupling has been previously investigated in two-population networks of oscillators~\cite{MA10}. In particular, it has been reported that coupling delay induces globally clustered chimera states in which the coherent and incoherent regions span both populations~\cite{SHE09b,SHE10}. Experimental evidence for chimera states in systems with time delay has been provided for chemical oscillators~\cite{TIN12} and electronic or optoelectronic systems~\cite{LAR13,LAR15}. Internal delayed feedback has been shown to induce chimeras in systems of globally coupled phase oscillators~\cite{YEL14} and laser networks \cite{BOE15}.
Chimera states in the presence of both delayed feedback and noise have been investigated in \cite{SEM15b}.

Here we investigate the interplay of noise and time-delayed feedback in a network of nonlocally coupled excitable elements and mainly focus on the role of feedback for coherence resonance chimeras. A distinctive feature of this type of chimera is that it is induced by noise and occurs in a certain restricted interval of noise intensity and systems parameters. The question we address here is whether these intervals can be increased by introducing time-delayed feedback. By exploring the impact of time delay, we uncover the mechanisms to control coherence resonance chimeras by time-delayed feedback. Our results show that applying time-delayed feedback promotes the occurrence of coherence resonance chimeras and induces new regimes.

\section{Model}\label{sec:system}

\begin{figure}[htbp]
\center{\includegraphics[width=1\linewidth]{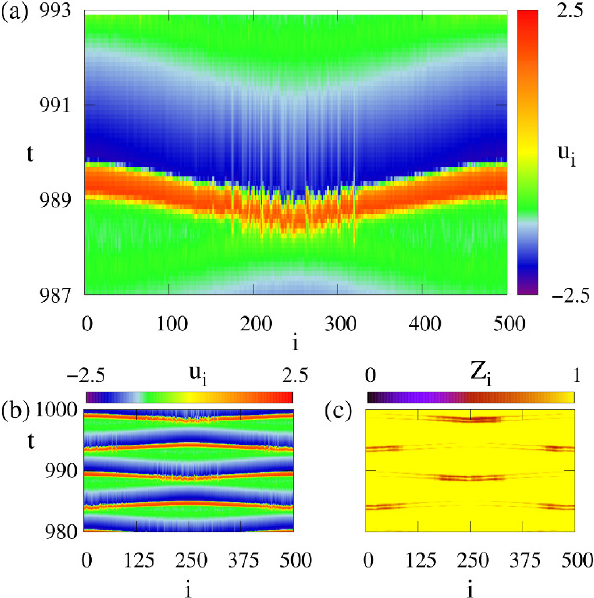}}
\caption[]{(a),(b) Space-time plots and (c) local order parameter for the coherence-resonance chimera. Initial conditions: randomly distributed on the circle $u^2 + v^2 = 4$. Parameters: $N=500$,  $\varepsilon=0.05$, $\phi=\pi/2-0.1$, $a=1.001$, $\sigma=0.4$, $r=0.2$, $D=0.0002$, $\gamma=0.2$, $\tau=1.0$.}
\label{fig:CR-chimera}
\end{figure}

We consider a ring of $N$ identical nonlocally coupled FitzHugh-Nagumo (FHN) systems with time-delayed feedback in the presence of Gaussian white noise:

\begin{equation}\label{eq:ring_fhn}
\begin{array}{c}
\varepsilon\frac{du_i}{dt}=u_i-\frac{u^3_i}{3}-v_i +\frac{\sigma}{2R}\sum\limits_{j=i-R}^{i+R} [b_{uu}(u_j-u_i)+\\ +b_{uv}(v_j-v_i)] + \gamma (u_i(t)-u_i(t-\tau)), \\
\frac{dv_i}{dt}=u_i+a+ \frac{\sigma}{2R}\sum\limits_{j=i-R}^{i+R} [b_{vu}(u_j-u_i)+\\ +b_{vv}(v_j-v_i)] + \sqrt{2D} \xi_{i}(t),
\end{array}
\end{equation}

where $u_i$ and $v_i$ are the activator and inhibitor variables, respectively, $i=1,...,N$ and all indices are modulo $N$, $\sigma$ is the coupling strength, $R$ is the number of nearest neighbours in each direction on a ring. We also introduce coupling range which is the normalized number of nearest neighbours $r=R/N$, where $N$ is the total number of elements in the network. Further, $\xi_i(t) \in \mathbb{R}$ is Gaussian white noise, i.e., $\langle \xi_i (t) \rangle \!=\! 0$ and $\langle \xi_i (t)  \xi_j(t') \rangle \!=\!  \delta_{ij} \delta(t-t'), ~\forall i,j$, and $D$ is the noise intensity. The feedback term is characterized by time delay $\tau$ and strength $\gamma$. A small parameter responsible for the time scale separation of fast activator and slow inhibitor is given by $\varepsilon>0$ and $a_i$ defines the excitability threshold. For an individual FHN element it determines whether the system is excitable ($|a_i|>1$), or oscillatory ($|a_i|<1$). In the present study we assume that all elements are in the excitable regime close to the threshold ($a_{i}\equiv a=1.001$ except for Figs.~\ref{fig:a_gamma=0,2}-\ref{fig:terms2}).
Eq.~(\ref{eq:ring_fhn}) contains not only direct, but also cross couplings between activator ($u$) and inhibitor ($v$) variables, which is modeled by a rotational coupling matrix \cite{OME13}:
\begin{equation}
B = \left(
\begin{array}{ccc}
b_{\mathrm{uu}} & & b_{\mathrm{uv}} \\
b_{\mathrm{vu}} & & b_{\mathrm{vv}}
\end{array}
\right) =
\left(
\begin{array}{ccc}
\cos \phi  & & \sin \phi \\
-\sin \phi  & & \cos \phi
\end{array}
\right),
\label{eq:Matrix_B}
\end{equation}
where $\phi\in[-\pi;\pi)$. Here we fix the parameter $\phi=\pi/2-0.1$. In the absence of time delay $\tau=0$ chimera states have been found for this value of $\phi$ in both the deterministic oscillatory  \cite{OME13} and the noisy excitable regime \cite{SEM16,ZAK17}. Moreover, it has been shown that chimera states occurring in the excitable regime \cite{SEM16,ZAK17} are different from those detected in the oscillatory regime~\cite{OME13}. In the presence of Gaussian white noise a special type of chimera state called {\em coherence resonance chimera} appears in a ring of $N$ nonlocally coupled excitable FHN systems (Fig.~\ref{fig:CR-chimera}).

In the present work, to control these patterns we introduce time-delayed feedback to the activator variable in Eqs.~(\ref{eq:ring_fhn}). For that purpose we fix all the parameters of the system in the regime of coherence resonance chimera and vary those characterizing the feedback term: $\gamma$ and $\tau$. For $\gamma=0$ Eqs. (\ref{eq:ring_fhn}) demonstrate coherence-resonance chimeras with the period $T\approx 4.76$. This regime can also be observed in the presence of time-delayed feedback for $\gamma=0.2$, $\tau=1.0$ and is shown as a space-time plot color-coded by the variable $u_{i}$ in Fig.~\ref{fig:CR-chimera}(a,b). One can clearly distinguish the regions of coherent and incoherent spiking.


To characterize spatial coherence and incoherence of chimera states one can use the local order parameter \cite{OME11,WOL11a}:
\begin{equation}
Z_k=\Big|\frac{1}{2\delta_Z}\sum\limits_{|j-k|\leq\delta_Z} e^{i \Theta_j}\Big|, \ \ \ k=1,\dots N
\end{equation}
where the geometric phase of the $j$-th element is defined by $\Theta_j=arctan(v_j/u_j)$ \cite{OME13} and $Z_k = 1$ and $Z_k<1$ indicate coherence and incoherence, respectively.
Figure~\ref{fig:CR-chimera}(c) represents a space-time plot color-coded by $Z_i$ and illustrates coexistence of coherent and incoherent domains with the latter characterized by values of $Z_i$ noticeably below unity (dark regions).

\begin{figure}[htbp]
\center{\includegraphics[width=1\linewidth]{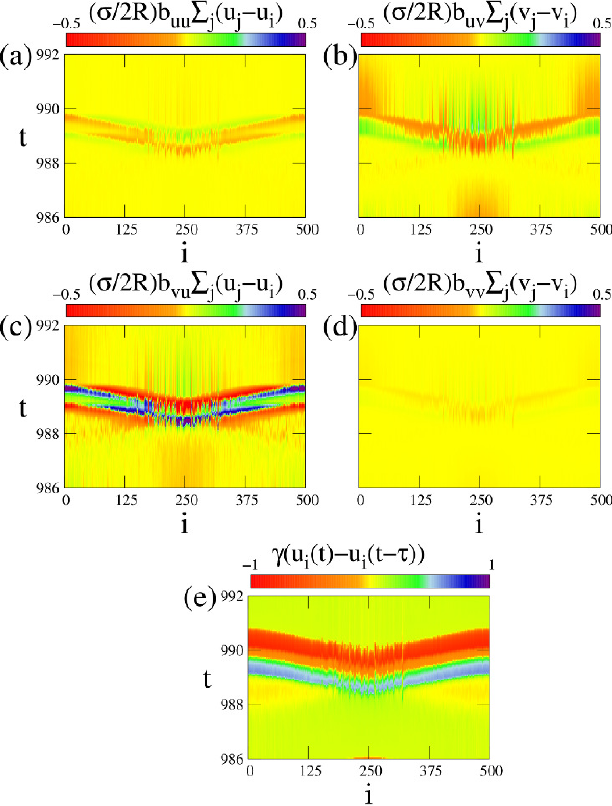}}
\caption[]{Space-time plots of coupling terms for $u$ and $v$ variables in the coherence-resonance chimera regime: (a) direct coupling for the $u$ variable, (b) cross-coupling for the $u$ variable, (c) cross-coupling for the $v$ variable, (d) direct coupling for the $v$ variable. (e) Space-time plot of the delay term. Parameters: $N=500$, $\varepsilon=0.05$, $\phi=\pi/2-0.1$, $a=1.001$, $\sigma=0.4$, $r=0.2$, $D=0.0002$, $\gamma=0.2$, $\tau=1.0$.}
\label{fig:coupl_func}
\end{figure}

One of the main features of these noise-induced chimera states is their alternating behavior which is absent in the oscillatory regime without noise. In more detail, the incoherent domain of the chimera pattern switches periodically its position on the ring, although its width remains fixed (Fig.~\ref{fig:CR-chimera}(b,c)). This property has been previously described in \cite{SEM16} and the explanation based on the time evolution of the coupling term has been provided in \cite{ZAK17}. Taking into account that the system~(\ref{eq:ring_fhn}) involves both direct and cross-couplings between activator $u$ and inhibitor $v$ variables, in total we have four coupling terms. It turns out that coupling terms form patterns shown as space-time plots in Fig.~\ref{fig:coupl_func}(a)--(d).


The crucial point is that the coupling acts as an additional term and shifts the nullclines of every individual element of the network. The coupling term with the strongest impact corresponds to cross-coupling for the variable $v$ (Fig.~\ref{fig:coupl_func}(c)). It means that the coupling significantly influences the $\dot{u}=0$ nullcline and shifts the threshold parameter $a$ which is responsible for the excitation. As a result for a certain group of nodes the threshold becomes lower due to coupling, and the probability of being excited by noise increases. Therefore, the elements of this group are the first to start the large excursion in the phase space and experience random spiking. The elements constituting the rest of the network spike coherently since they are pulled by already excited nodes and are, therefore, excited by coupling and not by noise. This scenario can also be obtained for the system Eq. (\ref{eq:ring_fhn}) in the presence of time-delayed feedback (Fig.~\ref{fig:coupl_func}). Due to the feedback an additional term appears in Eq.~(\ref{eq:ring_fhn}) and should be taken into account. Its evolution in time for all nodes of the network is shown in Fig.~\ref{fig:coupl_func}(e). The color-code bar clearly indicates that the values of the feedback term are larger than those of the coupling terms. However, for the chosen value of delay time $\tau=1.0$ the feedback does not have any essential impact on the behavior of coherence resonance chimeras since it is less than the intrinsic period of oscillations $T=4.76$ (Figs \ref{fig:CR-chimera}, \ref{fig:coupl_func}).

\begin{figure}[!h]
\begin{center}
\includegraphics[width=\linewidth]{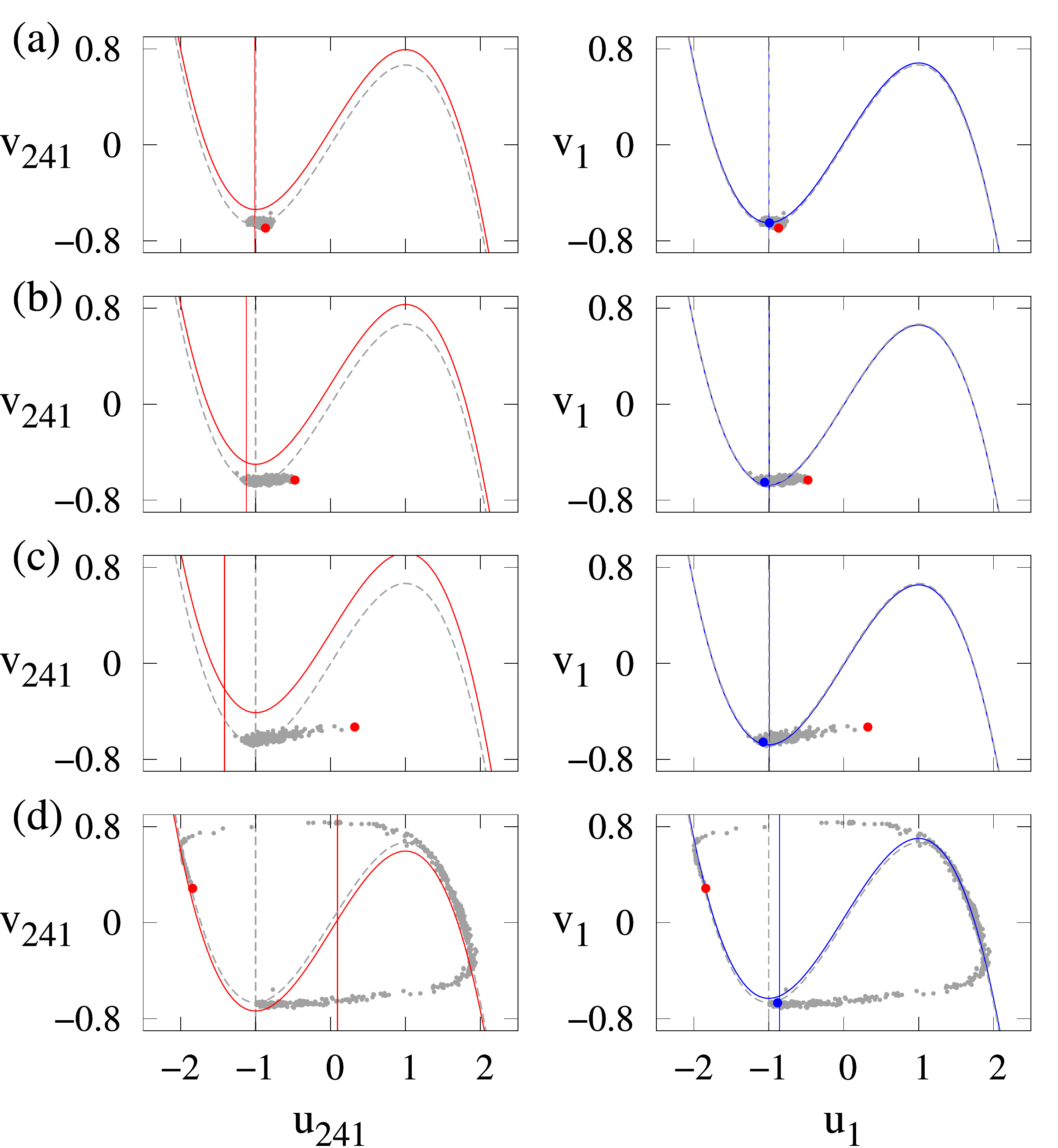}
\end{center}
\caption[]{Activator and inhibitor nullclines $\dot{u}$ and $\dot{v}$, respectively, for the selected nodes $i=241$ (left column) and $i=1$ (right column) of the system Eq.~(\ref{eq:ring_fhn}) in the coherence-resonance chimera regime for (a) $t=997.4$, (b) $t=997.6$, (c) $t=997.7$, (d) $t=998.4$. Parameters: $N=500$, $\varepsilon=0.05$, $a=1.001$, $\sigma=0.4$, $r=0.2$, $D=0.0002$, $\gamma=0.2$, $\tau=1.0$.}
\label{fig:nullcline}
\end{figure}

For the better understanding of this alternating dynamics in the presence of time-delayed feedback we study the impact of the coupling on activator and inhibitor nullclines for selected nodes of the system Eq.~(\ref{eq:ring_fhn}). In particular, we consider a sequence of phase portraits for the nodes $i=241$ (red dot) and $i=1$ (blue dot) which belong to the incoherent and coherent domains, respectively (Fig. \ref{fig:nullcline}). First, all the elements are located near the steady state (Fig. \ref{fig:nullcline}(a)). After a while the vertical nullcline of the node $i=241$ is shifted to the left of the value $u=-a=-1.001$ due to positive coupling term (panel (b)). Consequently, this node can be more easily excited by noise (panel (c)). Due to nonlocal coupling the excited node pulls its neighbours and they also start spiking. The coupling can also shift the vertical nullcline to the right of the value $u=-a=-1.001$ (panel (d)).

Similar results have been previously obtained for the case without time-delayed feedback \cite{ZAK17}. Therefore, for the strength $\gamma=0.2$ and time delay $\tau=1.0$ the feedback does not have an impact on the nullclines. Consequently, coherence resonance chimeras observed for small time delay of the feedback are the same as in the case without feedback.

\section{Dynamic regimes in the presence of time-delayed feedback}

\begin{figure}[htbp]
\center{\includegraphics[width=\linewidth]{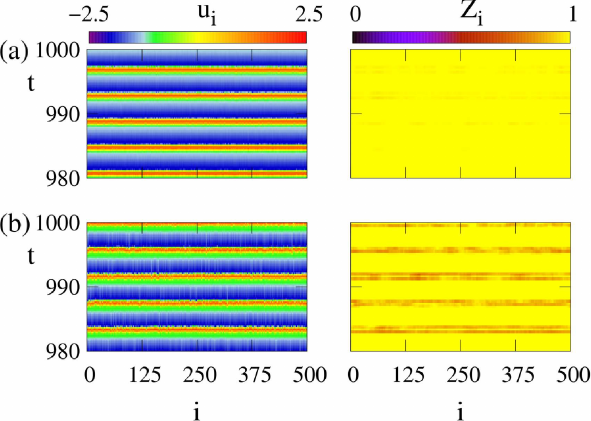}}
\caption[]{Space-time plots for the variable $u_i$ (left panels) and local order parameter $Z_i$ in the regime of (a) complete in-phase synchronization for $\gamma=0.4$, $\tau=3.6$ and (b) spatial incoherence for $\gamma=0.4$, $\tau=2.2$. Other parameters: $N=500$, $\varepsilon=0.05$, $a=1.001$, $\phi=\pi/2-0.1$, $D=0.0002$, $r=0.2$, $\sigma=0.4$.}
\label{fig:main_stp}
\end{figure}

Since our main goal is to study the impact of time-delayed feedback we now choose the parameters of the system in the regime of coherence resonance chimera and vary only the feedback parameters $\gamma$ and $\tau$. For fixed feedback strength $\gamma=0.4$ we observe the change of dynamic regimes by tuning the delay time $\tau$. For $\tau=3.6$ all the nodes of the network spike coherently, i.e, in-phase synchronization occurs (Fig.~\ref{fig:main_stp},a). The feedback with $\tau=2.2$ shifts the system into the regime which is incoherent in space and periodic in time: all the nodes demonstrate spiking behavior, but the spiking events of the neighboring nodes are not correlated (Fig.~\ref{fig:main_stp},b).

\begin{figure}[htbp]
\center{\includegraphics[width=1\linewidth]{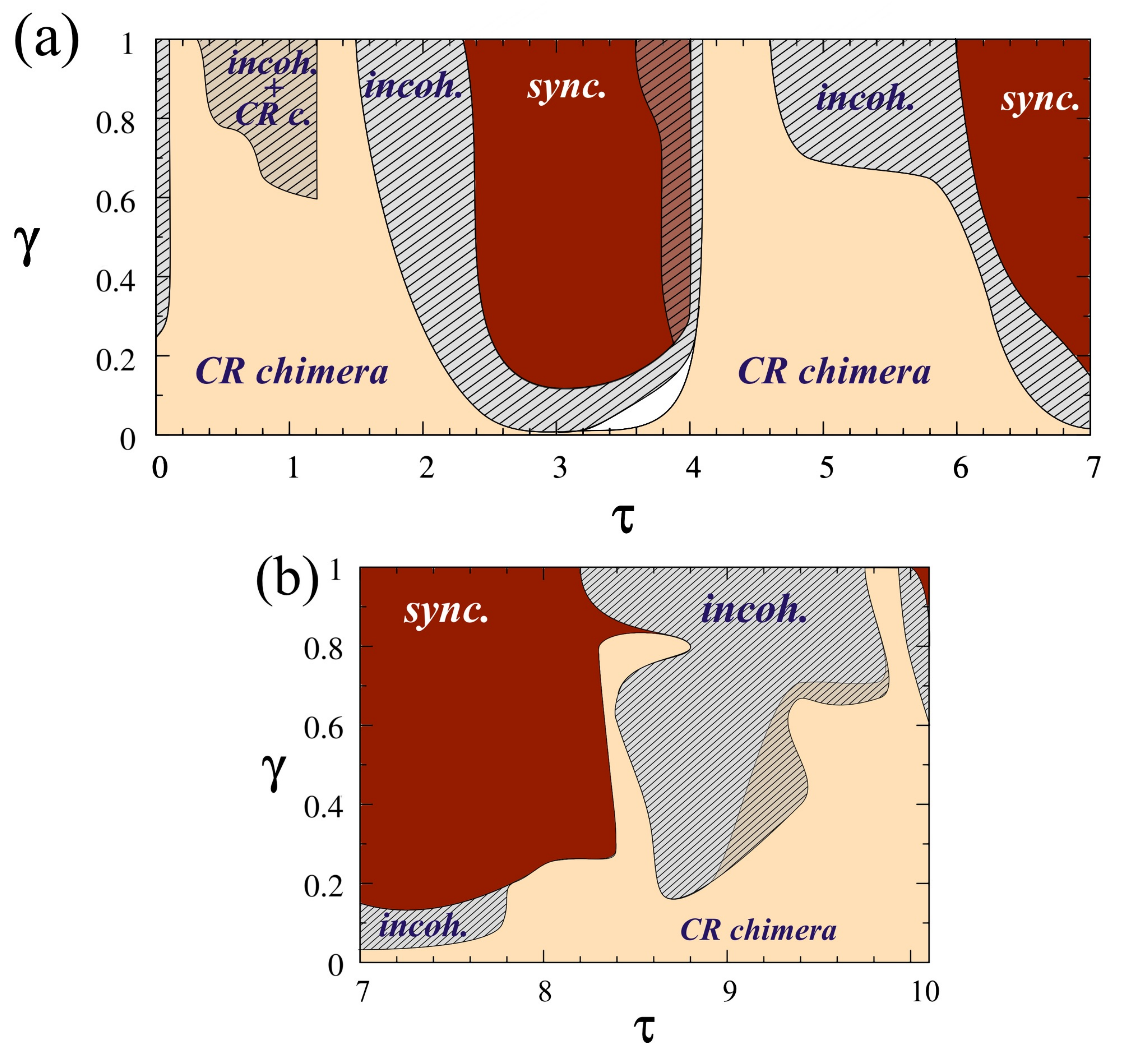}}
\caption[]{Dynamic regimes in the ($\tau,\gamma$) parameter plane. Red (dark-gray) regions: in-phase synchronization (see space-time plot in Fig.\ref{fig:main_stp}a); hatched regions: spatial incoherence (see space-time plot in Fig.\ref{fig:main_stp}b); white region: steady state; yellow (light-gray) regions: coherence resonance (CR) chimeras (see Fig.\ref{fig:main_stp}d). Parameters: $N=500$, $\varepsilon=0.05$, $a=1.001$, $\phi=\pi/2-0.1$, $D=0.0002$, $r=0.2$, $\sigma=0.4$.}
\label{fig:map}
\end{figure}

To gain a general view of the dynamics in the network of nonlocally coupled noisy excitable elements in the presence of time-delayed feedback we construct the map of regimes of the system Eq.(\ref{eq:ring_fhn}) in the ($\gamma$, $\tau$) parameter plane (Figure~\ref{fig:map}). For visualization reasons we have divided the map into two panels: panel (a) corresponds to the $\tau$ interval from $0$ to $7$ and includes the values $\tau\le T$, where $T$ is the period of the dynamics without delay ($T\approx 4.76$); panel (b) corresponds to larger values of $\tau$ including $\tau\approx 2T \approx 9.52$.

Note that the other parameters of the network are chosen in the coherence resonance chimera state which now occurs only for certain intervals of delay time $\tau$. We detect three main regions (yellow (light-grey) in Fig.~\ref{fig:map}) separated by in-phase synchronization domains (red (dark-grey) regions in Fig.~\ref{fig:map}) and regimes of spatially incoherent spiking (hatched regions Fig.~\ref{fig:map}). Although the map of regimes is dominated by various oscillatory patterns, for relatively small feedback strength $\gamma<0.2$ and time delay $3.2<\tau<4$ we also observe a small regime of steady state (white region in Fig.~\ref{fig:map}(a)).


Moreover, the diagram is characterized by multistability since spatially incoherent spiking can coexist with chimera states or in-phase synchronization. The overall structure of the map resembles a sequence of synchronization tongues although there are no clear resonances for delay times equal to the multiples of the intrinsic period $T\approx 4.76$. Nevertheless, applying the feedback with delay time $\tau\approx 2T \approx 9.52$ does not change the dynamics dramatically, and the regime of coherence resonance chimera is still observed for a wide range of feedback strength (Fig.~\ref{fig:map}).

\section{Impact of the feedback on coherence resonance chimera existence: noise intensity range}

Without feedback, as previously reported \cite{SEM16,ZAK17}, coherence resonance chimeras are observed for a certain restricted interval of noise intensity $0.000062\le D\le 0.000325$ for the following parameters of the system: $N=500$, $\varepsilon=0.05$, $a=1.001$, $\phi=\pi/2-0.1$, $r=0.2$, $\sigma=0.4$ (this set of parameters is fixed throughout this Section). Time-delayed feedback modifies this interval. To illustrate this effect we consider two cases: $\gamma<0.5$ and $\gamma>0.5$ which allows for a better understanding of the impact of feedback strength on this interval. Also for the two values of parameter $\gamma$ we choose different delay times $\tau$ from all three regions of coherence resonance chimeras shown in the ($\tau,\gamma$) plane in Fig.~\ref{fig:map}(a,b). Time-delayed feedback slightly changes the range of noise intensity values where chimera states occur in the system (\ref{eq:ring_fhn}) for both considered values of feedback strength: $\gamma=0.2$ (Fig.~\ref{fig:D_gamma=0,2}) and $\gamma=0.6$ (Fig.~\ref{fig:D_gamma=0,6}).


\begin{figure}[htbp]
\center{\includegraphics[width=\linewidth]{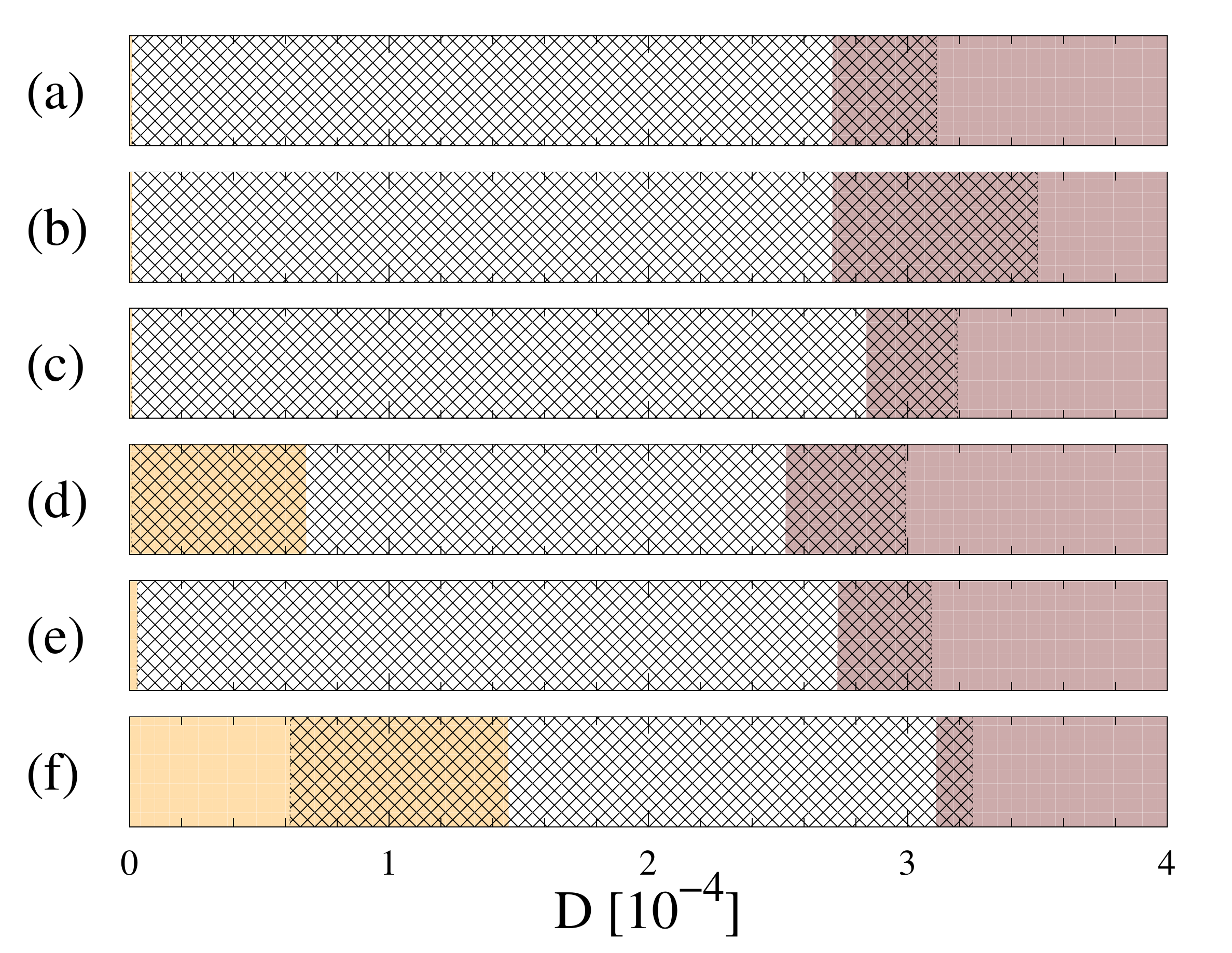}}
\caption[]{Dynamic regimes depending on the noise intensity $D$ for feedback strength $\gamma=0.2$ and different values of delay time: (a) $\tau=9.52$, (b) $\tau=6.0$, (c) $\tau=4.76$, (d) $\tau=1.8$, (e) $\tau=0.8$, (f) $\tau=0$. Dynamic regimes: steady state (yellow/light grey); spatially incoherent spiking (pink/dark grey); coherence resonance chimeras (hatching). Other parameters: $N=500$, $\varepsilon=0.05$, $a=1.001$, $\phi=\pi/2-0.1$, $r=0.2$, $\sigma=0.4$.}
\label{fig:D_gamma=0,2}
\end{figure}



\begin{figure}[htbp]
\center{\includegraphics[width=\linewidth]{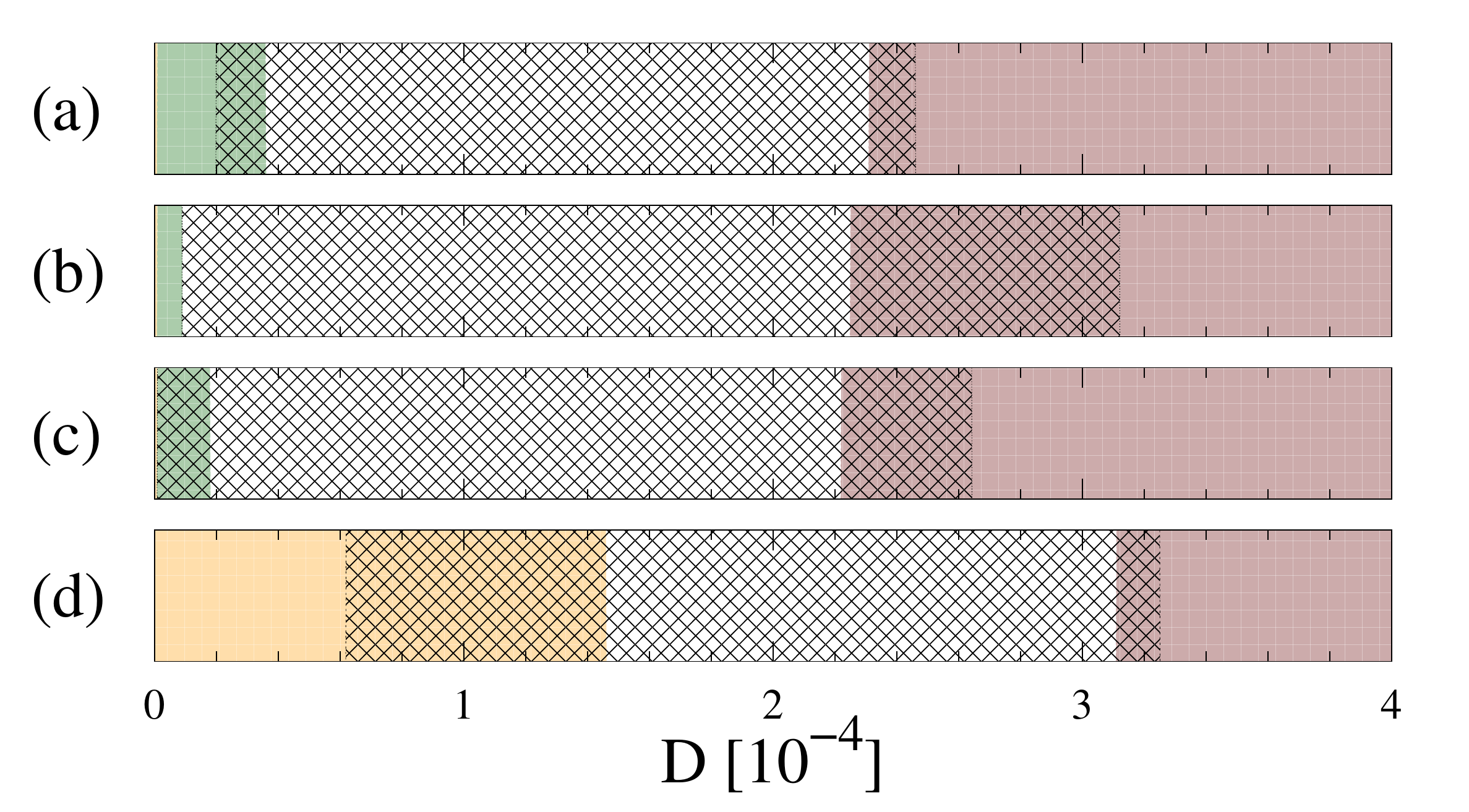}}
\caption[]{Dynamic regimes depending on the noise intensity $D$ for feedback strength $\gamma=0.6$ and different values of delay time: (a) $\tau=9.52$, (b) $\tau=4.76$, (c) $\tau=0.8$, (d) $\tau=0$. Dynamic regimes: steady state (yellow/light grey); spatially incoherent spiking (pink/dark-grey); synchronization (green/grey); coherence resonance chimeras (hatching). Other parameters: $N=500$, $\varepsilon=0.05$, $a=1.001$, $\phi=\pi/2-0.1$, $r=0.2$, $\sigma=0.4$.}
\label{fig:D_gamma=0,6}
\end{figure}


For rather weak feedback strength $\gamma=0.2$ the interval of existence of chimera patterns is enlarged for all the considered delay times. Interestingly, the right boundary of this interval can be shifted in the direction of stronger noise (Fig.~\ref{fig:D_gamma=0,2}b) as well as in the direction of lower noise intensities (Fig.~\ref{fig:D_gamma=0,2}a,d,e) and remains almost unchanged for delay time $\tau=4.76\approx T$ (Fig.~\ref{fig:D_gamma=0,2}c). Therefore, by appropriately choosing the feedback delay time one can adjust the value of noise intensity for which spatially incoherent spiking replaces coherence resonance chimeras within the interval $0.00030\le D\le 0.00035$ (Fig.~\ref{fig:D_gamma=0,2}). The transition from the steady state to coherence resonance chimeras for increasing noise occurs at the left boundary (Fig.~\ref{fig:D_gamma=0,2}f) which is shifted by the feedback to smaller noise intensities (Fig.~\ref{fig:D_gamma=0,2}e). Furthermore, due to feedback, chimera states appear even at zero noise intensity (Fig.~\ref{fig:D_gamma=0,2}(a)--(d)). Therefore, time-delayed feedback promotes coherence resonance chimeras not only by increasing the noise range where they exist, but also by inducing these patterns in the absence of noise. The largest range of $D$ corresponds to $\tau=6.0$ (Fig.~\ref{fig:D_gamma=0,2}b)
It is important to note that on the borders of the intervals the multistability is observed. Chimera states can coexist with the steady state on the left border and with the regime of spatially incoherent spiking on the right border.

Large feedback strength $\gamma=0.6$ can also shift the left boundary of the chimera interval to lower (Fig.~\ref{fig:D_gamma=0,6}a,b) and even zero (Fig.~\ref{fig:D_gamma=0,6}c) noise values. The multistability on the borders also occurs. Interestingly, for  $\gamma=0.6$the chimera state overlaps with the complete synchronization regime on the left boundary and not with the steady state as in the case of weak feedback strength. The right boundary strongly depends on $\tau$ and shifts into the direction of lower noise intensities (Fig.~\ref{fig:D_gamma=0,6}a,b,c). The largest detected interval for $\gamma=0.6$ corresponds to $\tau=4.76\approx T$ (Fig.~\ref{fig:D_gamma=0,2}b) and for $\tau=9.52\approx 2T$ we even observe shrinking of the interval (Fig.~\ref{fig:D_gamma=0,6}a).

If we compare the interval of chimera existence without time-delayed feedback $0.000062\le D\le 0.000325$ (Fig.~\ref{fig:D_gamma=0,2}f and Fig.~\ref{fig:D_gamma=0,6}d) with the interval the most enlarged by the feedback $0.000001\le D \le 0.00035$ it turns out that we achieve 33 per cent improvement rate.

\section{Impact of the feedback on coherence resonance chimera existence: threshold parameter range}

It has been previously shown that coherence resonance chimera can be obtained only in a small interval of $a$ ($0.995\le a\le 1.004$). To analyze the impact of time-delayed feedback we again consider two cases: $\gamma=0.2$ and $\gamma=0.6$ and different values of delay time. Figure~\ref{fig:a_gamma=0,2} corresponds to the case of small feedback strength $\gamma=0.2$ and Figure~\ref{fig:a_gamma=0,6} illustrates the results for the case of larger feedback strength $\gamma=0.6$.

\begin{figure}[htbp]
\center{\includegraphics[width=\linewidth]{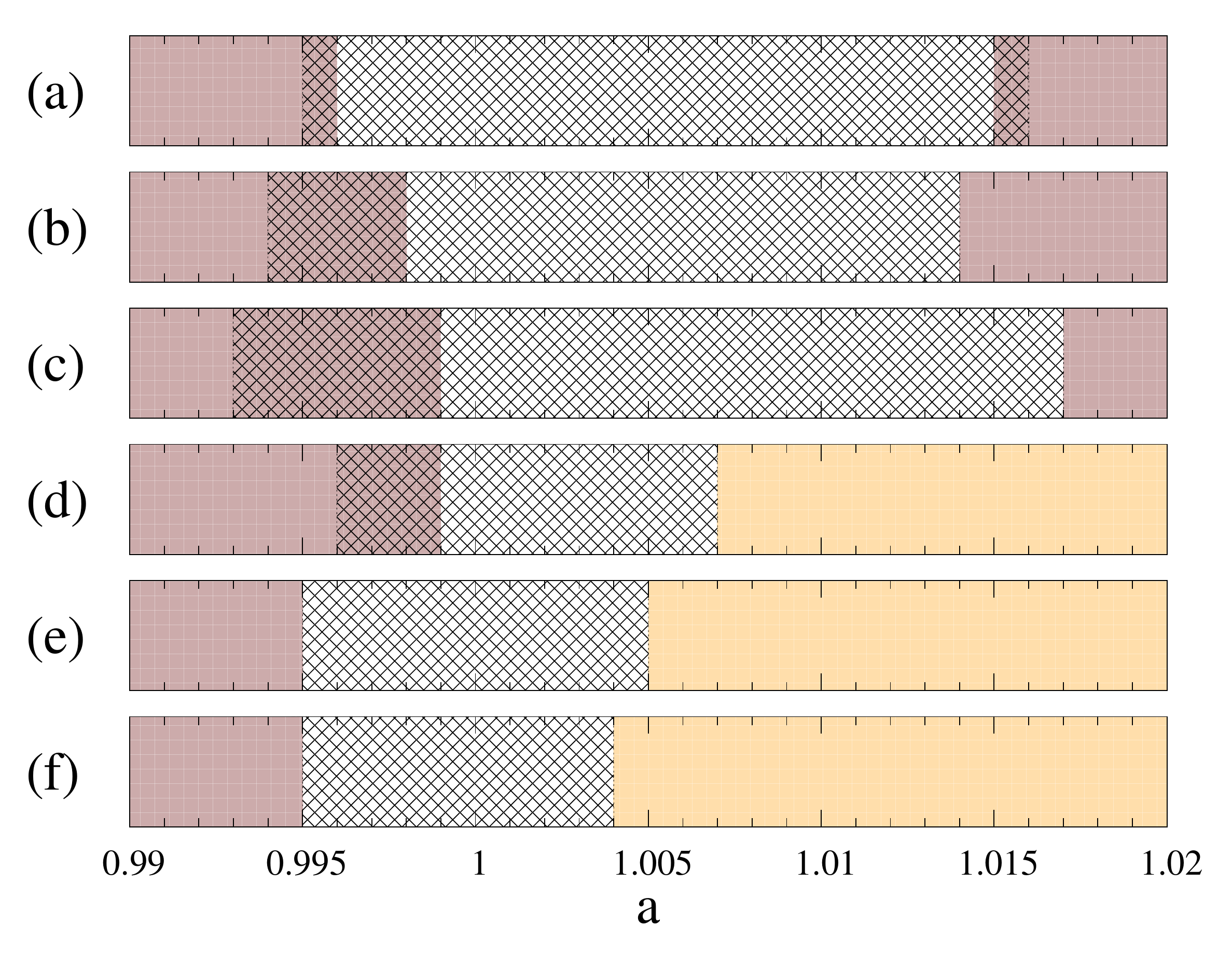}}
\caption[]{Dynamic regimes depending on the threshold parameter $a$ for feedback strength $\gamma=0.2$ and different values of delay time: (a) $\tau=9.52$, (b) $\tau=6.0$, (c) $\tau=4.76$, (d) $\tau=1.8$, (e) $\tau=0.8$, (f) $\tau=0$. Dynamic regimes: steady state (yellow/light grey); spatially incoherent spiking (pink/dark grey); coherence resonance chimeras (hatching). Other parameters: $N=500$, $\varepsilon=0.05$, $D=0.0002$, $\phi=\pi/2-0.1$, $r=0.2$, $\sigma=0.4$.}
\label{fig:a_gamma=0,2}
\end{figure}


\begin{figure}[htbp]
\center{\includegraphics[width=\linewidth]{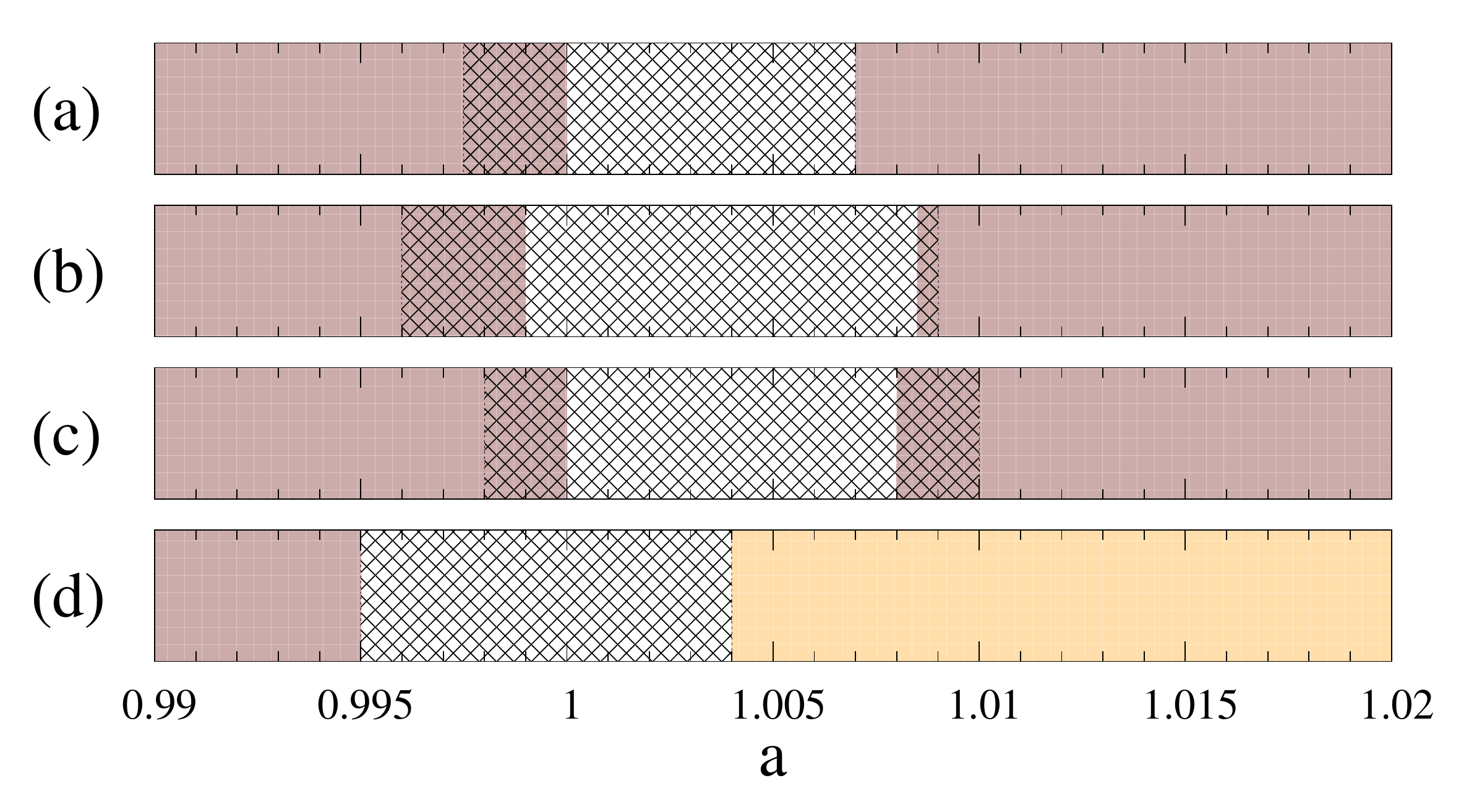}}
\caption[]{Dynamic regimes depending on the threshold parameter $a$ for feedback strength $\gamma=0.6$ and different values of delay time: (a) $\tau=9.52$, (b) $\tau=4.76$, (c) $\tau=0.8$, (d) $\tau=0$. Dynamic regimes: steady state (yellow/light grey); spatially incoherent spiking (pink/dark grey); coherence resonance chimeras (hatching). Other parameters: $N=500$, $\varepsilon=0.05$, $D=0.0002$, $\phi=\pi/2-0.1$, $r=0.2$, $\sigma=0.4$.}
\label{fig:a_gamma=0,6}
\end{figure}



For the two considered values of $\gamma$ time-delayed feedback significantly changes the range of the threshold parameter $a$ where coherence resonance chimeras exist. Moreover, in both cases this interval is increased the most when the delay time is equal to the intrinsic period of the system $\tau=4.76\approx T$ (Fig. \ref{fig:a_gamma=0,2}c and Fig. \ref{fig:a_gamma=0,6}b). However, smaller feedback strength allows for stronger enlargement of the interval: for $\gamma=0.2$ and $\tau=4.76$ it is $0.993\le a \le 1.017$ and is more than doubled compared to the case without feedback $0.995\le a\le 1.004$ (Fig. \ref{fig:a_gamma=0,2}c).

While tuning the threshold parameter $a$ we observe multistability on the boundaries of the coherence resonance chimera regime where this pattern coexists with spatially incoherent spiking (Fig. \ref{fig:a_gamma=0,2}a--d and Fig. \ref{fig:a_gamma=0,6}a--c). For increasing parameter $a$ the coherence resonance chimeras disappear in the absence of feedback, and a steady state is observed (Fig. \ref{fig:a_gamma=0,2}f, Fig. \ref{fig:a_gamma=0,6}d). However, for $\gamma=0.2$, $\tau\ge T$ (Fig. \ref{fig:a_gamma=0,2}a--c) and for all considered values of time delay in the case of strong feedback $\gamma=0.6$ (Fig. \ref{fig:a_gamma=0,6}a--c) the steady state is replaced by spatially incoherent spiking, i.e, the feedback induces oscillatory behaviour of the network.

\begin{figure}[htbp]
\center{\includegraphics[width=\linewidth]{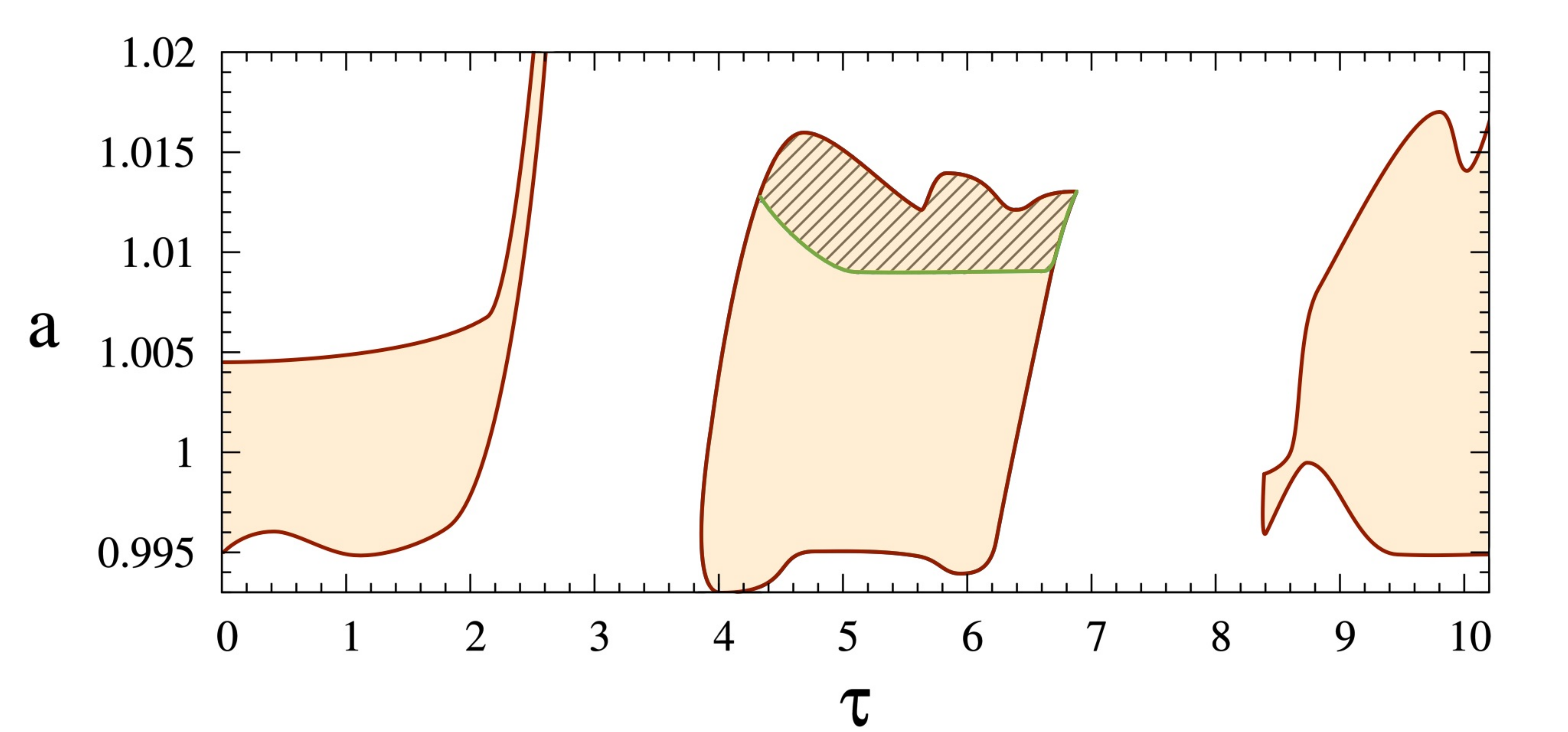}}
\caption[]{Coherence resonance chimera states in ($a,\tau$)-plane (yellow/grey regions). The hatched region corresponds to the period-two coherence resonance chimera. Parameters: $N=500$, $\varepsilon=0.05$, $D=0.0002$, $\phi=\pi/2-0.1$, $r=0.2$, $\sigma=0.4$, $\gamma=0.2$.}
\label{fig:a-tau}
\end{figure}

As it can be seen from Fig. \ref{fig:a_gamma=0,2}, for decreasing delay time $\tau$ from $9.52$ to $0$ we observe a nonlinear modulation of the size of the $a$-interval of existence of chimera states. To gain more insight into this effect, we define the parameter range for which this pattern exists in the ($a$, $\tau$) plane (Fig.~\ref{fig:a-tau}). We detect isolated regions occurring for certain disconnected intervals of $\tau$. The region centered at the time delay value close to the intrinsic period of the system $\tau = 4.76 \approx T$ clearly indicates the enlargement of the $a$-interval. 


\begin{figure}[htbp]
\center{\includegraphics[width=1\linewidth]{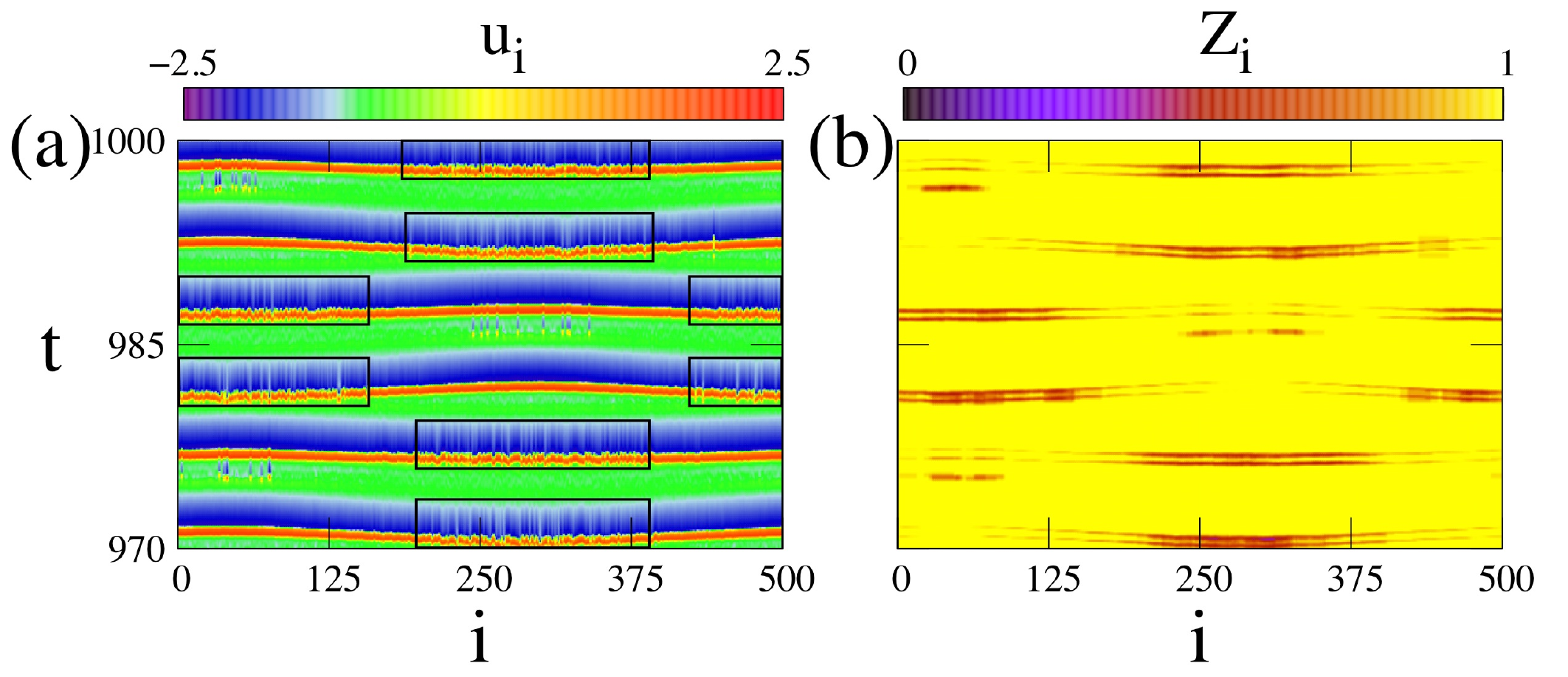}}
\caption[]{Space-time plot for the variable $u_i$ (a) and local order parameter $Z_i$ (b) in the regime of period-two coherence-resonance chimera. Initial conditions: randomly distributed on the circle $u^2 + v^2 = 4$. Incoherent domains are marked by rectangles in panel (a). Parameters: $N=500$, $\varepsilon=0.05$, $a=1.012$, $\sigma=0.4$, $r=0.2$, $D=0.0002$, $\gamma=0.2$, $\tau=4.76$.}
\label{fig:CR-chimera2}
\end{figure}

Interestingly, at the top of this region for $\tau\approx T$ and $a>1.01$ we find a novel chimera regime (hatching in Fig.~\ref{fig:a-tau}) which is induced by time-delayed feedback and has not been previously shown for the system (\ref{eq:ring_fhn}) without delay. The space-time plot for the variable $u_i$ and the local order parameter indicate the coexistence in space of coherent and incoherent spiking as well as alternating behavior, typical features of coherence resonance chimeras (Fig.~\ref{fig:CR-chimera2}). Furthermore, the alternation takes place periodically and the incoherent domain switches its position on the ring. However, the switching events occur not for every spiking cycle as in the coherence resonance chimera state (Fig.~\ref{fig:CR-chimera}b,c), but for every second spiking event (Fig.~\ref{fig:CR-chimera2}a,b). Due to this distinguishing feature we call this pattern \textit{period-two coherence resonance chimera}.


\begin{figure}[!h]
\begin{center}
\includegraphics[width=\linewidth]{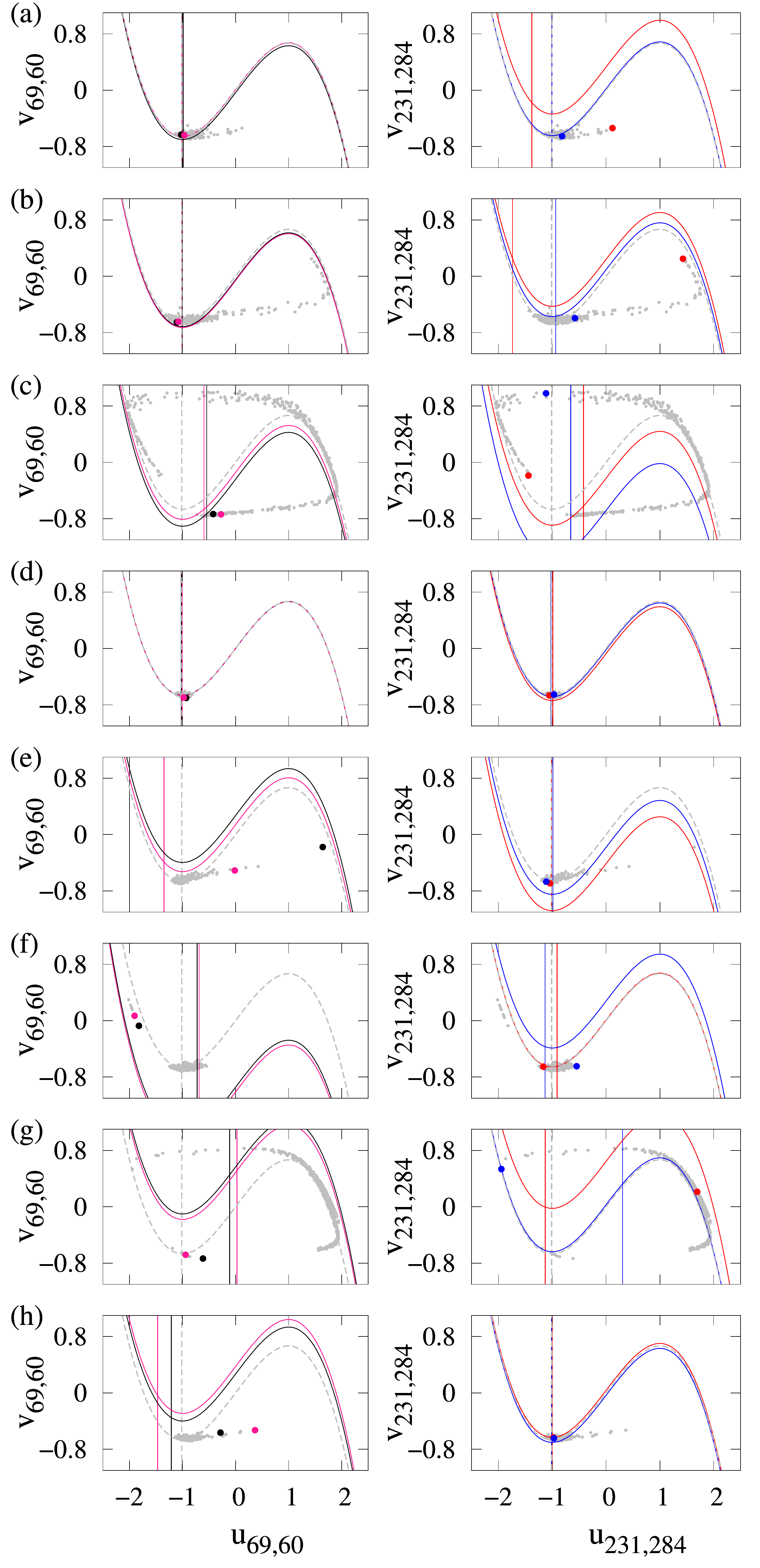}
\end{center}
\caption[]{Activator and inhibitor nullclines $\dot{u}$ and $\dot{v}$, respectively, for the selected nodes $i=69$ -- black, $i=60$ -- pink (left column) and $i=231$ -- red, $i=284$ -- blue (right column) of the system Eq.~(\ref{eq:ring_fhn}) in period-two coherence-resonance chimera regime for (a) $t=969.9$, (b) $t=970.15$, (c) $t=970.9$, (d) $t=974.5$, (e) $t=975.0$, (f) $t=975.9$, (g) $t=976.65$, (h) $t=980.55$. Parameters: $N=500$, $\varepsilon=0.05$, $a=1.012$, $\sigma=0.4$, $r=0.2$, $D=0.0002$, $\gamma=0.2$, $\tau=4.76$.}
\label{fig:nullcline2}
\end{figure}

To understand the mechanism of this alternation we consider a temporal sequence of the phase portraits of the system Eq. (\ref{eq:ring_fhn}) with the nullclines indicated for four selected nodes of the network. As it can be seen from Fig. \ref{fig:CR-chimera2} the incoherent domain alternates between two regions: the first region corresponds to the nodes $i\in [0,125],[450,500]$ and the second region is $i\in[200,375]$. For this reason we choose nodes $i=69$ and $i=60$ from the first region and nodes $i=231$ and $i=284$ from the second region. Next we analyze their dynamics during one period (Fig.~\ref{fig:nullcline2}).

We begin our observation when the incoherent spiking occurs in the second region $i\in[200,375]$. The node $i=231$ (red in Fig. \ref{fig:nullcline2} right column) starts the excursion in the phase space first since its nullclines are shifted and it can, therefore, be excited more easily by noise (right panel in Fig~\ref{fig:nullcline2}a). At the same time the elements $i=60$ and $i=69$ from the coherent domain (first region) rest in the steady state since their nullclines are unchanged (left panel in Fig~\ref{fig:nullcline2}a). 
Next, the nullclines of the other nodes from the second region are modified (see node $i=284$ (blue) in the right panel of Fig~\ref{fig:nullcline2}b,c). Consequently, they are now also excited by noise and, therefore, incoherently (right panel in Fig~\ref{fig:nullcline2}c), while the elements from the coherent domain still stay in the vicinity of the steady state with the nullclines unchanged (left panel in Fig~\ref{fig:nullcline2}c). Further, when the nodes from the incoherent domain are well on the way (right panel in Fig~\ref{fig:nullcline2}c) they pull the nodes from the first region that, therefore, also start spiking (left panel in Fig~\ref{fig:nullcline2}c). Since they are excited not by noise but due to the pulling of the neighbours their spiking is coherent.   
After performing a spike all the nodes return to the steady state (Fig~\ref{fig:nullcline2}d). This scenario is typical for coherence resonance chimeras (see Fig.~\ref{fig:nullcline}) for which the interchange of coherent and incoherent domains takes place during each subsequent excitation. However, this is not the case for period-two coherence resonance chimera as we see below.

Next we consider the second excitation for the elements of the network Eq. (\ref{eq:ring_fhn}). At the next moment in time a small group of nodes including $i=60$ and $i=69$ from the first region starts an excursion in the phase space due to noise (left panel in Fig~\ref{fig:nullcline2}e) while the elements from the second region remain in the steady state (right panel in Fig~\ref{fig:nullcline2}e). However, the spiking of this small group is weak since time-delayed feedback significantly shifts the nullclines (the cubic ones down and the vertical ones to the right) and does not allow the elements to make the full cycle in the phase space before going back to the steady state (left panel in Fig~\ref{fig:nullcline2}f). At the same time for the nodes from the second region the feedback shifts the cubic nullclines up and the vertical nullclines to the left making them more easily excitable by noise (right panel in Fig~\ref{fig:nullcline2}f). 
Therefore, the incoherent spiking is again induced in the second region while the nodes from the first region are pulled coherently due to coupling (Fig~\ref{fig:nullcline2}g). 
After that all nodes return again to the steady state. Next during the third excitation the nullclines for the elements from the first region are shifted in a way making the excitation threshold lower and, therefore, the spiking starts from the first region due to noise: the node $i=60$ is the first to spike (left panel in Fig~\ref{fig:nullcline2}h). Hence, finally the coherent and incoherent domains are interchanged and further the steps described above repeat with the only difference that the first region is now incoherent while the second corresponds to coherent spiking.

\begin{figure}[htbp]
\center{\includegraphics[width=1\linewidth]{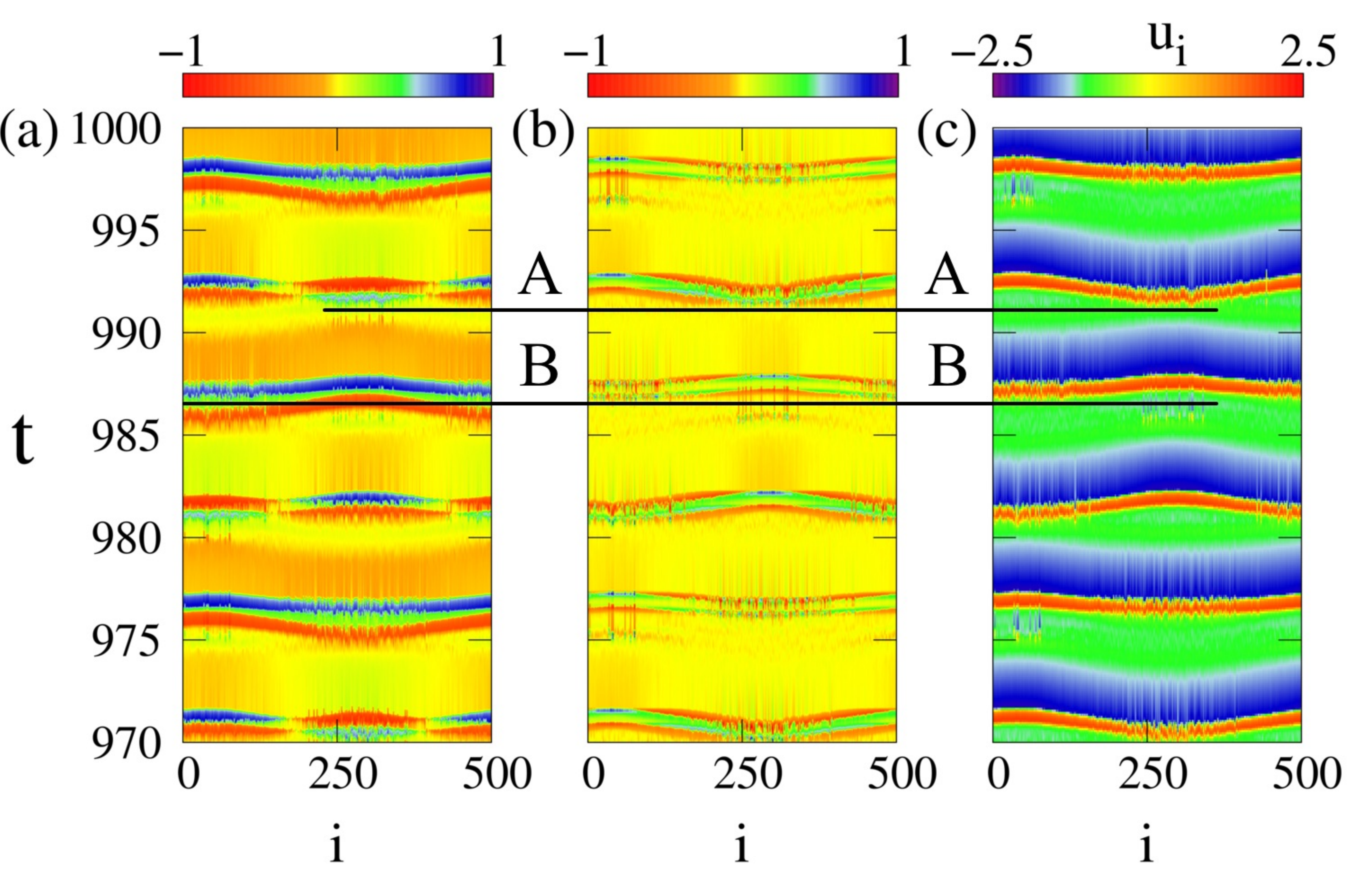}}
\caption[]{Space-time plots of coupling terms for $u$ and $v$ variables in the period-two coherence resonance chimera regime: (a) time-delayed feedback for the $u$ variable, (b) coupling for the $v$ variable. (c) Space-time plot of $u_i$ variable. Parameters: $N=500$, $\varepsilon=0.05$, $a=1.012$, $\sigma=0.4$, $r=0.2$, $D=0.0002$, $\gamma=0.2$, $\tau=4.76$.}
\label{fig:terms2}
\end{figure}

Thus, it is the time-delayed feedback that prevents the alternation for every spiking cycle. As it can be seen from Fig.~\ref{fig:coupl_func} the largest coupling term corresponds to cross-coupling in the $v$-equation of the system (\ref{eq:ring_fhn}). However, the contribution of the feedback term is significantly stronger than that of the coupling terms. For this reason alternating behaviour can only occur when the time-delayed feedback term is close to zero. To illustrate that we consider the impact of coupling terms and feedback upon the first and the second equation in system (\ref{eq:ring_fhn}) in the regime of two period coherence resonance chimera (Fig.~\ref{fig:terms2}). This figure clearly indicates that the interchange of coherent and incoherent domains in the chimera pattern occurs when the feedback term is close to zero (line A in Fig.~\ref{fig:terms2}). On the other hand, the alternation fails when the feedback term is non-zero and the coupling term almost vanishes (line B in Fig.~\ref{fig:terms2}).

\section{Conclusions}

In conclusion, we have investigated the impact of time-delayed feedback on the dynamics of a network of nonlocally coupled FitzHugh-Nagumo elements in the excitable regime in the presence of noise. Our special attention is given to a recently discovered chimera state, i.e., coherence resonance chimeras. We demonstrate that time-delayed feedback promotes this pattern: it allows for control of the range of parameter values where noise-induced chimera exists and in most cases increases this range. Moreover, the feedback induces coherence resonance chimeras for vanishing noise intensities. Additionally, we show that the threshold parameter interval of coherence resonance chimeras can be more than doubled by applying feedback with delay time close to the intrinsic period of the system. Compared to the case without feedback this provides an essential improvement which could be relevant for the experimental realization of coherence resonance chimeras. Furthermore, when the feedback delay coincides with the intrinsic period of the network we find a novel feedback-induced regime which we call period-two coherence resonance chimera. We explain the alternating behavior of this novel pattern by analyzing the evolution of the nullclines due to the coupling and feedback terms of the network. 

\section{Acknowledgments}
This work was supported by DFG in the framework of SFB 910 and by the Russian Ministry of Education and Science (project code 3.8616.2017) and the Russian Foundation for Basic Research (Grant No. 15-02-02288).


\end{document}